\documentclass[conference,9pt]{IEEEtran}

\makeatletter
\def\endthebibliography{%
	\def\@noitemerr{\@latex@warning{Empty `thebibliography' environment}}%
	\endlist
}
\makeatother

\usepackage{graphicx}
\usepackage{ifpdf}
\usepackage{epstopdf}
\usepackage{ifpdf}
\ifpdf
\DeclareGraphicsExtensions{.eps}
\else
\DeclareGraphicsExtensions{.eps}
\fi
\usepackage[noadjust]{cite}

\usepackage[cmex10]{amsmath}
\usepackage{amssymb}  
\usepackage{amsthm}

\newtheorem{proposition}{Proposition}

\newtheorem{lemma}{Lemma}

\IEEEoverridecommandlockouts

\newcommand\MYhyperrefoptions{bookmarks=true,bookmarksnumbered=true,
	pdfpagemode={UseOutlines},plainpages=false,pdfpagelabels=true,
	colorlinks=true,linkcolor={black},citecolor={black},urlcolor={black},
	pdftitle={WaterNetworks},
	pdfsubject={},
	pdfauthor={Mohammadhafez Bazrafshan and Nikolaos Gatsis},
	pdfkeywords={}}

\ifCLASSINFOpdf
\usepackage[\MYhyperrefoptions,pdftex]{hyperref}
\else
\usepackage[\MYhyperrefoptions,breaklinks=true,dvips]{hyperref}
\usepackage{breakurl}
\fi

\newcommand{\mr}[1]{\mathrm{#1}}
\newcommand{\mc}[1]{\mathcal{#1}}
\newcommand{\mb}[1]{\mathbf{#1}} 
\newcommand{\mbb}[1]{\mathbb{#1}}

\newcommand{\diag}{\mathrm{diag}}

 \usepackage{algpseudocode}
\usepackage{algorithm}
\newcommand{\eq}[2]{
	\begin{IEEEeqnarray}{#1}
		#2
\end{IEEEeqnarray}}

\begin{document}
	
	\title{A Fixed-Point Iteration for Steady-State Analysis\\ of Water Distribution Networks }
	\author{Mohammadhafez~Bazrafshan, Nikolaos~Gatsis, Marcio Giacomoni, and Ahmad Taha
		\thanks{M.~Bazrafshan is with the Department of Electrical and Computer Engineering, The Univ. of Texas at Austin. N.~Gatsis and A. Taha are with the Dept. of Electrical \& Computer Engineering,  The Univ. of Texas at San Antonio. M. Giacomoni is with the Department of Civil and Environmental Engineering, The Univ. of Texas at San Antonio. This material is based upon work partially supported by the National Science Foundation under Grants  CMMI-1728629 and CCF-1421583.
		}}
	\maketitle
	
	\begin{abstract}
		This paper develops a fixed-point iteration to solve the steady-state water flow equations in an urban water distribution network.  The fixed-point iteration is derived upon the assumption of turbulent flow solutions and the validity of the Hazen-Williams head loss formula for water flow. Local convergence is ensured  if the spectral radius of the Jacobian at the solution is smaller than one. The implication is that the solution is at least locally unique and that the spectral radius of the Jacobian provides an estimate of the convergence speed. A sample water network is provided to assert the application of the proposed method.
			\end{abstract}
	

\section{Introduction}
The steady-state  water flow is a fundamental problem in water distribution networks and amounts to solving for the water flow rates  in pipes and water pressures at nodes, given the rate of water consumption and delivery across the network nodes. Steady-state water flow analysis is required upon 	 water demand changes or upon network expansion to  ensure sufficient water pressures for satisfactory service. Furthermore, such an analysis also serves to evaluate optimality of  the procedures for  water network design~\cite{Vasan2010},  scheduling~\cite{Singh2018}, operations~\cite{Martinez2007, Mala-Jetmarova2017},  control~\cite{Ormsbee1994}, as well as joint optimization of water and energy networks in smart cities~\cite{Zamzam2017,Li2018}. 

The water flow problem involves  solving a set of nonlinear equalities in an equal number of unknown variables.  The unknown variables comprise water flow rates in each pipe and the total head at each node. The latter  serves as a proxy for pressure. The equalities are derived based on applying the momentum equation, the continuity equation, and the energy equation.  The momentum equation describes the nonlinear relationship between head loss and the water flow rate in a pipe and is typically determined experimentally. The continuity equation  ensures conservation of water flow rate at a node, and  the energy equation  states that the head loss is  equal to the difference of total head between the two ends of a pipe~\cite{ChinWater2006}. 

 Traditionally, three methods are used to compute the solution to the steady-state water flow problem~\cite{Jeppson1974}:  Hardy Cross, Newton-Raphson, and the  Linear Theory Method.  The Hardy Cross method~\cite{Cross1936} was popularized in the early stages due to its simplicity. Upon an initial guess satisfying the continuity equation, the Hardy Cross method iteratively finds an approximate correction factor for flow rates by using Taylor expansion and accounting for the fact that the sum of head loss in a loop  amounts to zero. The iterations continue until the corrections in flow rates are sufficiently small. 
The Newton-Raphson method promises fast convergence upon provision of a good starting point, but it requires the computation of the inverse Jacobian per iteration. The  efficient gradient formulation of~\cite{Todini1987} which is the core computational engine for the steady-state flow analysis of the  water simulation software EPANET~\cite{EPANETManual} is based on Newton-Raphson. 
The Linear Theory Method \cite{WoodCharles1972} uses the value of water flow rate in a previous iteration to linearize the nonlinear momentum formula and recomputes the value of water flow rate in a new iteration. The connection of the Linear Theory Method as an approximate Newton-Raphson has been explicitly stated in~\cite[Section~4.3]{StephensonBook1984}. One of its main advantages over the Newton-Raphson and the Hardy Cross method is that it typically does not require a good initialization point for flow rates~\cite[Ch.3]{Jeppson1974}.

Inspired by the renewed interest in fixed-point methods for the traditional power flow problem in electrical networks~\cite{Bolognani2016,WangBernstein2018,BazGatsis-ZBus,BernsteinWangDallAnese2018}, we set out to investigate the application of a fixed-point iteration for solving the water flow equations.  The advantage is that one can then leverage the rich theory of contraction mappings to pursue conditions for local or global convergence and uniqueness of solutions in steady-state analysis; all of which have historically been recognized as crucial in confirming reliability of mathematical models for water distribution networks~\cite{Boulos1993},~\cite{Todini2008}.  Furthermore, reports suggest that fixed-point type methods may provide convergence even when the de-facto software fails to do so~\cite{Zhang2017a}.  The analysis of the fixed-point method in this paper relies on a condition for local convergence that also estimates the convergence speed, at least towards the end of the algorithm.

This paper is organized as follows.  Section~\ref{Sec:NetModel} introduces the network model, the nonlinear momentum equation for head loss and water flow, the continuity equation, the energy equation,  and finally formulates the water flow problem.  The fixed-point iteration is presented in Section~\ref{Sec:FixedPoint} to along with a  condition for this algorithm to be a local contraction. Section~\ref{Sec:NumTests} applies the fixed-point iteration to a sample distribution network and verifies the contraction condition. The paper concludes in Section~\ref{Sec:Conclusion} with pointers to future work.

\section{Network model and the water flow problem}
\label{Sec:NetModel}
This section presents the network model pertaining to steady-state analysis of water distribution networks, that is, network quantities represent values for a single snap-shot or time period. Let us denote a water distribution network with a directed graph $(\mc{N},\mc{L})$ where $\mc{N}=\{0,\ldots,N\}$ is the set of $N+1$ nodes and $\mc{L} \subseteq \mc{N} \times \mc{N}=\{1,\ldots,L\}$ is the set of  $L$ links. If link $\ell$ corresponds to the unordered tuple $(i,j)$  in the graph, we assume a direction for $\ell$ from $\min\{i,j\} \rightarrow \max\{i,j\}$. Nodes and links represent physical components in the network and are explained in what follows. 

Nodes comprise  junctions, reservoirs, and tanks. Junctions are nodes that consume water, reservoirs are infinite sources or sinks of water while tanks  can consume or inject water with a limited capacity~\cite{EPANETManual}. We assume that node $0 \in \mc{N}$ corresponds to a main reservoir, while the remaining nodes are indexed within $\mc{N}_+=\{1,\ldots, N\}$.     
Quantities of interest for nodes $n \in \mc{N}$ are the rate of water injection, denoted by $s_n$, and the hydraulic head, denoted by $h_n$.   Since reservoirs are sources of water, it conventionally holds that $s_n \ge 0$, while for junctions we have that $s_n<0$.  The hydraulic head, $h_n$, acts a proxy for water pressure.  The water injection rates and hydraulic heads are respectively collected in vectors $s=\{s_n\}_{n \in \mc{N}_+}$ and $h=\{h_n\}_{n \in \mc{N}_+}$, and define further  $s_{\mc{N}}=[s_0,\;  s']'$ and $h_{\mc{N}}=[h_0, \; h']'$,  where $(.)'$ denotes transposition. 

Links represent pipes, pumps, and control valves. This papers focuses on networks with pipes, and other elements will be included in future work. The quantities of interest for pipe $\ell \in \mc{L}$ are the rate of water flow, denoted by $q_\ell$, as well as the head loss, denoted by $\hbar_\ell$.   The head loss for pipe $\ell \in \mc{L}$, which serves as a proxy for pressure drop across the pipe,   is related to the rate of water flow on pipe $\ell \in \mc{L}$ through a \emph{momentum} equation.  Assuming the customary U.S.  units,  that is, head loss measured in feet and rate of water flow in cubic feet per second, a commonly used  head loss formula for turbulent flow is the Hazen-Williams equation, as follows:
\eq{rCl}{
\hbar_\ell:=\hbar_{\ell}(q_{\ell})=A_{\ell}|q_{\ell}|^{0.852} q_{\ell}, \quad \ell \in \mc{L} \label{Eq:HeadLossHazen}
}
where $A_{\ell}=4.727C_{\ell}^{-1.852}d_{\ell}^{-4.871}l_{\ell}$; $d_{\ell}$ and $l_{\ell}$ are respectively the diameter and length of a circular pipe $\ell$ measured in feet, and $C_{\ell}$ is a unitless coefficient, called the Hazen-Williams roughness coefficient. For new pipes, the value of  $C_{\ell}$ is typically above $100$.  The notation $\hbar_{\ell}(.)$ denotes a functional dependence of the head loss $\hbar_{\ell}$ to the flow $q_{\ell}$.  The flow rates and head losses are respectively collected in vectors $q=\{q_{\ell}\}_{\ell \in \mc{L}}$ and $\hbar=\{h_{\ell}\}_{\ell \in \mc{L}}$.  Furthermore, define $\hbar(q)=\{h_{\ell}(q_{\ell})\}_{\ell \in \mc{L}}$. 

Two main equations govern the steady-state behavior of water networks: the \emph{continuity} equation and the \emph{energy} equation.  The continuity equation, which is analogous to KCL in electrical networks,  states that the rate of water injection  into node $n \in \mc{N}$ equals the total rate of water flowing out on the links connected to node $n$.  

Using graph theory, the continuity equation can be mathematically expressed as follows:
\eq{rCl}{
s_{\mc{N}}&=&\mc{I}_{\mc{N}} q  \label{Eq:Continuity}
}
where $\mc{I}_{\mc{N}} \in \mbb{R}^{N+1} \times \mbb{R}^{L}$ is the graph incidence matrix defined as
\eq{rCl}{
\left[\mc{I}_{\mc{N}}\right]_{i,\ell}=\begin{cases}
	+1, & \text{if $\ell$ is directed out of } i\\
	-1,  & \text{if $\ell$ is directed into } i.
\end{cases} \label{Eq:Incidence}
}

The energy equation states that total head at the upstream node of the pipeline is equal to the total head at the downstream node of the pipeline plus any head losses occurring on the way.  The energy equation is expressed  as follows:
\eq{rCl}{
\hbar(q)=\mc{I}_{\mc{N}}' h_{\mc{N}}. \label{Eq:Energy}
}

Recall from graph theory that the vector of all ones, $\mb{1}_{N+1}$, is in the nullspace of $\mc{I}_{\mc{N}}'$.  Consider the partition of the incidence matrix as $\mc{I}_{\mc{N}}=[ \mc{I}_0, \; \mc{I}']'$ where $\mc{I}_0'$ denotes the row of $\mc{I}_{N}$ corresponding to the reservoir node~$0$, and $\mc{I}$ accounts for the remaining nodes of $\mc{I}_N$. Then, we have that
\eq{rCl}{
\mc{I}_{\mc{N}}' \mb{1}_{N+1} = \mc{I}_0+\mc{I}'\mb{1}_{N}=\mb{0}_{L}. \label{Eq:I0INDerive}
}
Thus, it holds that
\eq{rCl}{
\mc{I}_0&=&-\mc{I}' \mb{1}_{N} \label{Eq:I0inIN}
}

Using~\eqref{Eq:I0inIN} in~\eqref{Eq:Continuity} and~\eqref{Eq:Energy} we arrive at the water flow equations:
\eq{rCl}{\label{EqGroup:WaterFlow} \IEEEyesnumber \IEEEyessubnumber*
s&=& \mc{I}q, \label{Eq:WaterFlowContinuity} \\
\hbar(q)&=&\mc{I}'(h-h_0\mb{1}_N ), \label{Eq:WaterFlowEnergy}
}
Given the total reference head at the main reservoir $h_0$, and the vector of injections $s \in \mbb{R}^{N}$, the goal of the water flow problem~\eqref{EqGroup:WaterFlow} is to determine  the flow rates on all links, that is $q \in \mbb{R}^{L}$, and the total head at all remaining nodes, that is $h \in \mbb{R}^{N}$. Notice that the number of unknowns are equal to $N+L$ and so is the number of equations since~\eqref{Eq:WaterFlowContinuity} has $N$ entries and~\eqref{Eq:WaterFlowEnergy} has $L$ entries. Due to the nonlinearity of the left hand side of \eqref{Eq:WaterFlowEnergy},  a Jacobian based iterative method is typically applied; see e.g., \cite[Appendix D]{EPANETManual}.  Upon solving~\eqref{EqGroup:WaterFlow}, the flow rates solution $q^*$ determines the amount of water flow intake from the main reservoir, $s_0^*$:
\eq{rCl}{
s_0^*&=& \mc{I}_0q^*. \label{Eq:s0Compute}
}

\section{Fixed-point iteration}\label{Sec:FixedPoint}
This section is concerned with the development of a fixed-point method to solve the water-flow problem and its local convergence analysis.  It is assumed that there is a minimum flow rate level $q_{\min}>0$ such that the magnitude of all flow rates in the network are above that level. This assumption is consistent with postulating that the Reynolds number corresponding to all flows in the network are above a certain value. The Reynolds number characterizes the pattern of the flow in each pipe, and is related to the flow rate, per the following:
\eq{rCl}{
\mr{Re}=\frac{d_\ell |q_\ell|}{v S_\ell}
\label{Eq:Reynolds}
}
where  $d_\ell$ was defined previously as the diameter of the pipe, $S_\ell$ is the cross-sectional area of the pipe, and $v$ is the kinematic viscosity of the water. For example, if the network flows are turbulent, then the Reynolds number is greater than $4,000$, which corresponds to a minimum flow rate through~\eqref{Eq:Reynolds}.

%

The previous assumption enables to rewrite~\eqref{Eq:HeadLossHazen} as follows:
\eq{rCl}{
q&=& {A}^{-1} \diag(|q|^{-0.852}) \hbar \label{Eq:qhbar}
}
where $|q|^{-0.852}$ is a vector with entries $q_\ell^{-0.852}$ for $\ell \in \mc{L}$, and  $A=\diag(\{A_{\ell}\}_{\ell \in \mc{L}})$, where $\diag(.)$ represents a matrix whose off-diagonals are zero and its diagonals are populated with the vector $(.)$.
Replacing $q$ in \eqref{Eq:WaterFlowContinuity} by its equivalent in~\eqref{Eq:qhbar} yields
\eq{rCl}{
s&=& \mc{I} A^{-1} \diag(|q|^{-0.852}) \hbar  
}
Using the right hand side of~\eqref{Eq:WaterFlowEnergy} to replace $\hbar$ in the latter yields
\eq{rCl}{
s&=& \mc{I} A^{-1} \diag(|q|^{-0.852}) \mc{I}'(h-h_0\mb{1}_N)  \label{Eq:sII'}
}

The matrix  $\mc{I} A^{-1} \diag(|q|^{-0.852}) \mc{I}'$ is indeed invertible; as asserted by the following lemma.
\begin{lemma} \label{Lemma:Invertible}
	In a connected network, with flow rates above a minimum level, $\mc{I} A^{-1} \diag(|q|^{-0.852}) \mc{I}'$ is invertible.
\end{lemma}
\begin{IEEEproof}
For a connected network, $\mc{I}$ is full row rank. In addition, the matrix $G=A^{-1} \diag(|q|^{-0.852})$ is diagonal with positive entries on the diagonal. Therefore, the matrix $\mc{I} G \mc{I}'$ is the weighted Laplacian of the network and is positive definite~\cite[Ch.~13]{GodsilRoyle}. As such, it is also invertible. 
\end{IEEEproof}

Using the previous lemma, it follows from~\eqref{Eq:sII'} that
\eq{rCl}{ \label{Eq:h-h0s}
(h-h_0\mb{1}_N)&=& \left[\mc{I} A^{-1} \diag(|q|^{-0.852}) \mc{I}'\right]^{-1} s. \IEEEeqnarraynumspace
}
Multiplying both sides of~\eqref{Eq:h-h0s} by $\mc{I}'$ and invoking~\eqref{Eq:WaterFlowEnergy} yields
\begin{equation}
\hbar = \mc{I}'\left[\mc{I} A^{-1} \diag(|q|^{-0.852}) \mc{I}'\right]^{-1} s. 
\end{equation}
Invoking the latter into~\eqref{Eq:qhbar} yields a fixed-point map for $q$: 
\eq{rCl}{
	q&=& T(q) \label{Eq:FixedPointMap} 
}
where $T(q)$ equals the following expression in terms of $q$:
\eq{l}{
A^{-1} \diag(|q|^{-0.852}) \mc{I}'\left[\mc{I} A^{-1} \diag(|q|^{-0.852}) \mc{I}'\right]^{-1} s. \label{Eq:Tq} \IEEEeqnarraynumspace
}

It is worth emphasizing that~\eqref{Eq:FixedPointMap} is a set of equations for the flow rates $q$ that satisfy the water flow equations~\eqref{EqGroup:WaterFlow}. In other words, any solution $q^*$ that satisfies~\eqref{Eq:FixedPointMap} also satisfies~\eqref{EqGroup:WaterFlow} and vice versa. If a solution $q^*$ of~\eqref{Eq:FixedPointMap} is available, then the head losses can be computed by~\eqref{Eq:HeadLossHazen}, and the heads using~\eqref{Eq:h-h0s}. 

Using the fixed-point map in~\eqref{Eq:FixedPointMap}, an iterative method to solve the water flow problem~\ref{EqGroup:WaterFlow} indexed by $k=1,2,3,\ldots$ and initialized by $q^0$ can be constructed as follows:  
\begin{equation}
q^{k+1} = T(q^k).
\label{Eq:FixedPointIteration}
\end{equation}
Algorithm~\ref{Algorithm:FixedPointIteration} summarizes the steps.  The last step relies on~\eqref{Eq:h-h0s}. 
\begin{algorithm}[t] 
	\caption{Solve for $q,h$ in water-flow problem~\eqref{EqGroup:WaterFlow} \label{Algorithm:FixedPointIteration}}
	\begin{algorithmic}[1]
		\State Initialize turbulent flow rate $q^0$, i.e., $\forall \ell, q_\ell \ge q_{\min}$
		\State  $k\leftarrow 0$
		\While{$\|q^k-T(q^k)\|_{\infty}>\epsilon$} 
		\State $q^{k+1} \leftarrow T(q^k)$
		\State $k \leftarrow k+1$
		\EndWhile 
		\State $q^* \leftarrow q^k$ 
		\State $h^*=\left[\mc{I} A^{-1} \diag(|q^*|^{-0.852}) \mc{I}'\right]^{-1} s+h_0\mb{1}_N$ \label{AlgorithmLastStep}
	\end{algorithmic}
\end{algorithm}

The convergence of~\eqref{Eq:FixedPointIteration} depends on the Jacobian matrix of $T(q)$, denoted by $J(q)=\frac{\partial{T(q)}}{\partial q}$.  In particular, the Jacobian of $T(q)$ can be obtained using first-order Taylor approximation arguments similar to~\cite[pp.~644]{Boyd2004}, and its expression is provided by the next lemma.
\begin{lemma} \label{Lemma:Jacobian}
	The Jacobian matrix $J(q)$ of the map $T(q)$ is given by
\eq{rCl}{
J(q)=A^{-1}\left[ F +E \mc{I}'Z^{-1}\mc{I}H \right]\diag\left(\mc{I}'Z^{-1}s\right) \IEEEeqnarraynumspace \label{Eq:JacobianComputed}
}
where 
\eq{rCl}{\label{EqGroup:JacobConstants} \IEEEyesnumber \IEEEyessubnumber*
E&=&\diag(|q|^{-0.852}) \label{Eq:E}\\
F&=&-0.852\diag(|q|^{-1.852})\diag(\mr{sign}(q))  \label{Eq:F}\\
H&=& A^{-1}	\diag(|q|^{-1.852})\diag(\mr{sign}(q))  \label{Eq:H}\\
Z&=& \mc{I} A^{-1} E \mc{I}'.  \label{Eq:Z} 
}
\end{lemma}

The following proposition provides a condition for the local convergence of~\eqref{Eq:FixedPointIteration}.
\begin{proposition} \label{Proposition:LocalContraction}
Suppose that $q^*$ is a fixed-point of the map in~\eqref{Eq:Tq}, that is, $q^*=T(q^*)$.  Let $J^*=\frac{\partial{T(q)}}{\partial q}|_{q=q^*}$ be the Jacobian of the map $T(q)$ evaluated at $q^*$. Denote by $\lambda_i(J^*)$ the eigenvalues of $J^*$ for $i \in \mc{L}$  and define the spectral radius of $J^*$ as $\rho(J^*)=\max_{i}\{|\lambda_i(J^*)|\}$. If $\rho(J^*)<1$, then $T(q)$ is locally a contraction map around $q^*$, and $q^*$ is a locally unique fixed point.
	\end{proposition}
\begin{IEEEproof}
In view of the expression in~\eqref{Eq:JacobianComputed}, the entries of the Jacobian matrix (partial derivatives) are continuous at $q^*$. This fact together with the spectral radius condition enable us to invoke the Ostrowski Theorem, which yields the desired results~\cite[Sec.~10.1]{OrtegaRheiboldt}. 
\end{IEEEproof}	
	
The consequence of the previous proposition asserts that if all eigenvalues of $J^*$ have magnitude less than one, then the iterative method~\eqref{Eq:FixedPointIteration} converges to $q^*$ if initialized in a neighborhood around $q^*$. In addition, the solution is unique in this neighborhood. Finally, the contraction property characterizes the speed of convergence; in particular the distance between successive iterates decreases by a factor $\alpha\in(0,1)$: 
\begin{equation}
\|q^{k+1}-q^{k}\|_{\infty} \leq \alpha {\|q^{k}-q^{k-1}\|_{\infty}}
\label{Eq:ConvRate}
\end{equation}
In fact, the value of $\alpha$ is roughly $\rho(J^*)$~\cite[Sec.~10.1]{OrtegaRheiboldt}.
It should be noted however that Proposition~\ref{Proposition:LocalContraction} does not characterize the size of the neighborhood around $q^*$ where the previous results hold. 

In the next section, we test the proposed fixed-point iteration in solving the water flow problem~\eqref{EqGroup:WaterFlow} for a sample network.
\section{Numerical tests}
\label{Sec:NumTests}
\begin{figure}
	\centering
\includegraphics[scale=0.3]{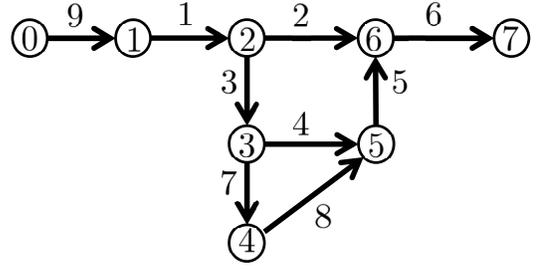}
\caption{Schematic of the example network.  Node $0$ is the main reservoir with $h_0=850$ feet. Assumed flow direction of link $\ell$ is from the node with a lower index to a node with higher index. }
\label{Fig:SampleNetwork}
\end{figure}

The network under study is a modified and simplified version of the example network from \cite[Ch.~2]{EPANETManual}.  A schematic is provided in Fig.~\ref{Fig:SampleNetwork} with pipe parameters given in Table~\ref{Table:PipeLine}. Node $0$ is a designated reservoir with  $h_0=850$ feet.  The vector of demands is  $s=[0, -150, -150,-200, -150, 0, -300]'$ in Gallons per minute and the negative sign denotes the consumption. 

\begin{figure}[t]
	\centering 
	\includegraphics[scale=0.4]{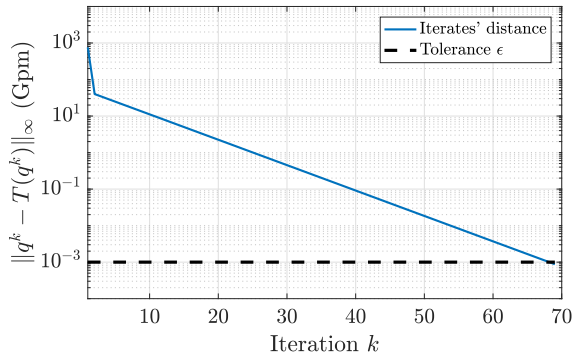}
\caption{Convergence of the fixed point iteration.  The difference between the value of $q^k$ and the mapping $T(q^k)$ approaches zero. }
\label{Fig:Convergence}
\end{figure}

\begin{figure}[t]
	\centering 
	\includegraphics[scale=0.4]{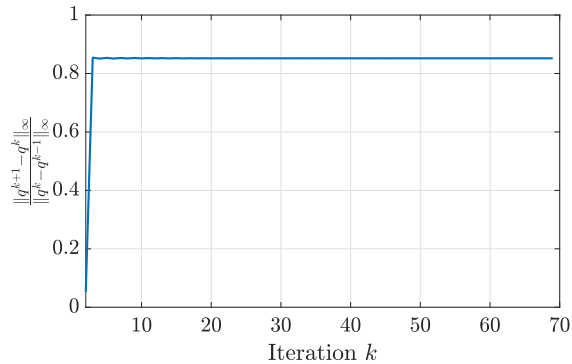}
	\caption{Convergence rate of the fixed point iteration.  The ratio $\frac{\|q^{k+1}-q^{k}\|_{\infty}}{\|q^{k}-q^{k-1}\|_{\infty}}$ is shown for $k=2,\ldots,k^*$. It seems that after just a few iterations, the sequence proceeds similar to a geometric sequence with a common ratio of $0.85$. }
	\label{Fig:Ratio}
\end{figure}

\begin{table}
	\centering 
	\label{Table:PipeLine}
	\caption{Pipe parameters}
	\renewcommand{\arraystretch}{1.2}
	\begin{tabular}{c|c|c|c}
		Pipe No. & Length (ft.) & Diameter (in.) &  Hazen-Williams $C$ \\
		\hline \hline
		1 & 3000 & 14 & 100 \\
		2 & 5000 & 12 & 100 \\
		3 & 5000 & 8 & 100 \\
		4 & 5000 & 8 & 100 \\
		5 & 5000 & 8 & 100\\
		6 & 7000 & 10 & 100 \\
		7 & 5000 & 6 & 100 \\
		8 & 7000 & 6 & 100 \\
		9 &3000 & 14 & 100\\
		\hline
		\end{tabular}
\end{table}
For Algorithm~\ref{Algorithm:FixedPointIteration}, the vector of initial water flows $q^0= 600\mb{1}_{L}$ is selected. The value of $\epsilon$ is set to $0.001$  Gallons per minute, which is a quite aggressive accuracy requirement. The algorithm takes $k^*=69$ iterations to achieve the desired tolerance. The solution for water flow (rounded to up to two decimals) and the total head values are computed to be
\eq{rCl}{\IEEEyesnumber \label{EqGroup:Solution} \notag  
	q^*&=&[815.03, 446.65, 218.38, 3.35, -146.65, 
	 300.00, \\ &&  65.03, -134.97, 815.03]' \text{ Gallons per minute }  \IEEEyessubnumber* \\
	h^*&=&[846.01, 842.01,  833.14, 829.32,  833.14, \notag \\
	&&   837.38, 829.84]' \: \text{feet}.} 
We  crosschecked the values in~\eqref{EqGroup:Solution} with a Jacobian-based nonlinear solver, namely MATLAB's \texttt{fsolve}. The maximum difference between the solutions in~\eqref{EqGroup:Solution} and the ones computed by \texttt{fsolve} are on the order of $10^{-4}$ Gallons per minute for water flow and on the order of $10^{-10}$ feet for total head.   Fig.~\ref{Fig:Convergence} depicts the progression of $\|q^{k+1}-q^{k}\|_{\infty}$ on a logarithmic scale per iteration $k$, and is shown to decrease linearly with the iteration index $k$. 
In Fig.~\ref{Fig:Ratio}, the rate of convergence, that is, the ratio $\frac{\|q^{k+1}-q^{k}\|_{\infty}}{\|q^{k}-q^{k-1}\|_{\infty}}$,  is shown [cf.~\eqref{Eq:ConvRate}].  It turns out that after a few iterations, the sequence $q^k$ proceeds according to a geometric progression with common ratio of $0.85$. 
Last but not least, we evaluate the condition of Proposition~\ref{Proposition:LocalContraction}.  
Upon computing the Jacobian using~\eqref{Eq:JacobianComputed}, the spectral radius was found to be $\rho(J^*)=0.8520<1$. This validates the hypothesis that the fixed point map is locally a contraction, and the spectral radius is surprisingly close to the ratio of distances between successive iterates provided by Fig.~\ref{Fig:Ratio}. 
\section{Concluding remarks and future directions}
\label{Sec:Conclusion}

Leveraging ideas from graph theory, this paper develops a fixed-point method to solve the steady-state water flow problem, which amounts to a set of nonlinear equations relating the flow rates in the network with the heads at junctions. The Jacobian of the fixed-point map is used to shed light in the convergence properties  of the algorithm, including the speed of convergence, at least locally. 

The focus of this paper is on networks where all links are pipes. It is worth enlarging the scope of the algorithm to include other types of links, such as pumps and control valves, as well as other types of nodes, including tanks or emitters, whose water outflow rates depends on the pressure. Furthermore, the head loss along a pipe was modeled after the Hazen-Williams equation, while more accurate but involved expressions may be used~\cite{ChinWater2006}.

A further direction is towards more sophisticated analysis of the fixed-point map, which can potentially lead to sufficient conditions for global contraction. The significance is that a unique solution to the water flow equations then exists over a larger region of flow rates, while algorithm convergence is established even if the initialization is not close to the solution. Indeed, contraction mapping approaches have been successful in demonstrating the convergence of traditional algorithms for the solution of the power flow problem in power networks, as well as for the development of sufficient conditions for the existence and uniqueness of the power flow solution in single-phase~\cite{Bolognani2016, WangBernstein2018} and multi-phase distribution networks~\cite{BazGatsis-ZBus}.

\bibliographystyle{IEEEtran}
\bibliography{water}

\begin{thebibliography}{10}
\providecommand{\url}[1]{#1}
\csname url@samestyle\endcsname
\providecommand{\newblock}{\relax}
\providecommand{\bibinfo}[2]{#2}
\providecommand{\BIBentrySTDinterwordspacing}{\spaceskip=0pt\relax}
\providecommand{\BIBentryALTinterwordstretchfactor}{4}
\providecommand{\BIBentryALTinterwordspacing}{\spaceskip=\fontdimen2\font plus
\BIBentryALTinterwordstretchfactor\fontdimen3\font minus
  \fontdimen4\font\relax}
\providecommand{\BIBforeignlanguage}[2]{{%
\expandafter\ifx\csname l@#1\endcsname\relax
\typeout{** WARNING: IEEEtran.bst: No hyphenation pattern has been}%
\typeout{** loaded for the language `#1'. Using the pattern for}%
\typeout{** the default language instead.}%
\else
\language=\csname l@#1\endcsname
\fi
#2}}
\providecommand{\BIBdecl}{\relax}
\BIBdecl

\bibitem{Vasan2010}
A.~Vasan and S.~P. Simonovic, ``{Optimization of Water Distribution Network
  Design Using Differential Evolution},'' \emph{J. Water Resour. Plan. Manag.},
  vol. 136, no.~2, pp. 279--287, Mar 2010.

\bibitem{Singh2018}
\BIBentryALTinterwordspacing
M.~K. Singh and V.~Kekatos, ``{Optimal Scheduling of Water Distribution
  Systems},'' Jun 2018. [Online]. Available:
  \url{http://arxiv.org/abs/1806.07988}
\BIBentrySTDinterwordspacing

\bibitem{Martinez2007}
F.~Mart{\'{i}}nez, V.~{Hern{\'{a}} Ndez}, M.~Alonso, Z.~Rao, S.~Alvisi, and
  V.~{Hern{\'{a}} Ndez Jos{\'{e}}}, ``{Optimizing the operation of the Valencia
  water-distribution network},'' \emph{Journal of Hydroinformatics}, no.~1, pp.
  65--78, Jan. 2007.

\bibitem{Mala-Jetmarova2017}
H.~Mala-Jetmarova, N.~Sultanova, and D.~Savic, ``{Lost in optimisation of water
  distribution systems? A literature review of system operation},''
  \emph{Environ. Model. Softw.}, vol.~93, pp. 209--254, Jul. 2017.

\bibitem{Ormsbee1994}
L.~E. Ormsbee and K.~E. Lansey, ``{Optimal Control of Water Supply Pumping
  Systems},'' \emph{J. Water Resour. Plann. Manage.}, vol. 120, no.~2, pp.
  237--252, Mar. 1994.

\bibitem{Zamzam2017}
\BIBentryALTinterwordspacing
A.~S. Zamzam, E.~Dall'Anese, C.~Zhao, J.~A. Taylor, and N.~D. Sidiropoulos,
  ``{Optimal Water-Power Flow Problem: Formulation and Distributed Optimal
  Solution},'' Aug. 2017. [Online]. Available:
  \url{http://arxiv.org/abs/1708.06754}
\BIBentrySTDinterwordspacing

\bibitem{Li2018}
\BIBentryALTinterwordspacing
Q.~Li, S.~Yu, A.~S. Al-Sumaiti, and K.~Turitsyn, ``{Micro Water-Energy Nexus:
  Optimal Demand-Side Management and Quasi-Convex Hull Relaxation},'' May 2018.
  [Online]. Available: \url{http://arxiv.org/abs/1805.07626}
\BIBentrySTDinterwordspacing

\bibitem{ChinWater2006}
D.~Chin, \emph{Water-resources engineering}.\hskip 1em plus 0.5em minus
  0.4em\relax Upper Saddle River, N.J: Pearson Prentice Hall, 2006.

\bibitem{Jeppson1974}
``{Steady Flow Analysis of Pipe Networks: An Instructional Manual},''
  \emph{Reports. Pap.}, vol. 300, 1974.

\bibitem{Cross1936}
H.~Cross, ``{ANALYSIS OF FLOW IN NETWORKS OF CONDUITS OR CONDUCTORS},''
  \emph{Engineering Experiment Station}, no. 286, Nov. 1936.

\bibitem{Todini1987}
E.~Todini and S.~Pilati, ``{A gradient method for the analysis of pipe
  networks},'' in \emph{Int. Conf. Comput. Appl. Water Supply Distrib.
  Leicester Polytech. UK}, no. August, 1987.

\bibitem{EPANETManual}
L.~a. Rossman, ``{EPANET 2: users manual},'' \emph{Cincinnati US Environ. Prot.
  Agency Natl. Risk Manag. Res. Lab.}, vol.~38, no. Sept., 2000.

\bibitem{WoodCharles1972}
D.~J. Wood and C.~O.~A. Charles., ``{Hydraulic Network Analysis Using Linear
  Theory},'' \emph{Jour. of the Hydraulics Div., ASCE}, vol.~98, pp.
  1157--1170, Jul. 1972.

\bibitem{StephensonBook1984}
D.~Stephenson, \emph{Pipeflow Analysis}.\hskip 1em plus 0.5em minus 0.4em\relax
  New York, NY: Elsevier, 1984.

\bibitem{Bolognani2016}
S.~Bolognani and S.~Zampieri, ``{On the Existence and Linear Approximation of
  the Power Flow Solution in Power Distribution Networks},'' \emph{IEEE Trans.
  Power Syst.}, vol.~31, no.~1, pp. 163--172, Jan. 2016.

\bibitem{WangBernstein2018}
C.~Wang, A.~Bernstein, J.~Y.~L. Boudec, and M.~Paolone, ``Explicit conditions
  on existence and uniqueness of load-flow solutions in distribution
  networks,'' \emph{IEEE Trans. Smart Grid}, vol.~9, no.~2, pp. 953--962, March
  2018.

\bibitem{BazGatsis-ZBus}
M.~Bazrafshan and N.~Gatsis, ``Convergence of the {Z}-bus method for
  three-phase distribution load-flow with {ZIP} loads,'' \emph{IEEE Trans.
  Power Syst.}, vol.~33, no.~1, pp. 153--165, Jan. 2018.

\bibitem{BernsteinWangDallAnese2018}
\BIBentryALTinterwordspacing
A.~Bernstein, C.~Wang, E.~Dall'Anese, J.-Y.~L. Boudec, and C.~Zhao,
  ``{Load-Flow in Multiphase Distribution Networks: Existence, Uniqueness, and
  Linear Models},'' Feb. 2017. [Online]. Available:
  \url{http://arxiv.org/abs/1702.03310}
\BIBentrySTDinterwordspacing

\bibitem{Boulos1993}
P.~F. Boulos, T.~Altman, and J.~C.~P. Liou, ``{On the solvability of water
  distribution networks with unknown pipe characteristics},'' \emph{Appl. Math.
  Modelling}, vol.~17, pp. 380--387, Jul. 1993.

\bibitem{Todini2008}
E.~Todini, ``{On The Convergence Properties of the Different Pipe Network
  Algorithms},'' in \emph{Water Distrib. Syst. Anal. Symp. 2006}.\hskip 1em
  plus 0.5em minus 0.4em\relax Reston, VA: American Society of Civil Engineers,
  mar 2008, pp. 1--16.

\bibitem{Zhang2017a}
\BIBentryALTinterwordspacing
H.~Zhang, X.~Cheng, T.~Huang, H.~Cong, and J.~Xu, ``{Hydraulic Analysis of
  Water Distribution Systems Based on Fixed Point Iteration Method},''
  \emph{Water Resour. Manag.}, vol.~31, no.~5, pp. 1605--1618, Mar. 2017.
  [Online]. Available: \url{http://link.springer.com/10.1007/s11269-017-1601-1}
\BIBentrySTDinterwordspacing

\bibitem{GodsilRoyle}
C.~Godsil and G.~Royle, \emph{Algebraic Graph Theory}.\hskip 1em plus 0.5em
  minus 0.4em\relax New York, NY: Springer, 2001.

\bibitem{Boyd2004}
\BIBentryALTinterwordspacing
S.~Boyd and L.~Vandenberghe, \emph{{Convex Optimization}}.\hskip 1em plus 0.5em
  minus 0.4em\relax Cambridge University Press, 2004. [Online]. Available:
  \url{https://web.stanford.edu/{~}boyd/cvxbook/bv{\_}cvxbook.pdf}
\BIBentrySTDinterwordspacing

\bibitem{OrtegaRheiboldt}
J.~M. Ortega and W.~C. Rheinboldt, \emph{Iterative Solution of Nonlinear
  Equations in Several Variables}.\hskip 1em plus 0.5em minus 0.4em\relax New
  York, NY: Academic Press, 1970.

\end{thebibliography}

\end{document}